\documentstyle[pra,aps,twocolumn,epsf]{revtex}

\begin{document}
\draft
\title{Experimental realization of Popper's Experiment:\\
Violation of the Uncertainty Principle?}
\author{Yoon-Ho Kim\thanks{Email: yokim@umbc.edu} and Yanhua Shih\thanks{Email: shih@umbc.edu}}
\address{Department of Physics\\
University of Maryland, Baltimore County\\ Baltimore, Maryland
21250, U.S.A.}
\date{Revised 19 October 1999, to appear in Foundations of Physics}
\maketitle

\begin{abstract}
An entangled pair of photons (1 and 2) are emitted to opposite directions. A
narrow slit is placed in the path of photon 1 to provide the precise knowledge
of its position on the $y$-axis and this also determines the precise
$y$-position of its twin, photon 2, due to quantum entanglement.  Is photon 2
going to experience a greater uncertainty in momentum, that is, a greater $
\Delta p_{y}$ because of the precise knowledge of its position $y$? The
experimental data show $\Delta y\Delta p_{y}<h $ for photon 2. Can this recent
realization of the thought experiment of Karl Popper signal a violation of the
uncertainty principle?
\end{abstract}
\pacs{}

\section{Introduction}

Uncertainty, one of the basic principles of quantum mechanics, distinguishes
the world of quantum phenomena from the realm of classical physics. Quantum
mechanically, one can never expect to measure both the precise position and
momentum of a particle at the same time. It is prohibited. We say that the
quantum observables ``position'' and ``momentum'' are ``complementary'' because
the precise knowledge of the position (momentum) implies that all possible
outcomes of measuring the momentum (position) are equally probable.

Karl Popper, being a ``metaphysical realist", however took a different point of
view. In his opinion, the quantum formalism {\em could} and {\em should} be
interpreted realistically: a particle must have precise position and momentum,
which shares the same view as Einstein. In this regard he invented a thought
experiment in the early 1930's which aimed to support the realistic
interpretation of quantum mechanics and undermine the Copenhagen
orthodoxy\cite{Popper1}. What Popper intends to show in his thought experiment
is that a particle can have both precise position and momentum at the same time
through the correlation measurement of an entangled two-particle system.  This
bears striking similarity to what EPR {\em gedankenexperiment} of 1935 seeks to
conclude \cite{EPR}. But different from EPR's {\em gedankenexperiment}, the
physics community remained ignorant of Popper's experiment.

In this paper we wish to report a recent realization of Popper's thought
experiment. Indeed, it is astonishing to see that the experimental results
agree with Popper's prediction. Through quantum entanglement one may learn the
precise knowledge of a photon's position and would therefore expect a greater
uncertainty in its momentum under the usual Copenhagen interpretation of the
uncertainty relations. However, the measurement shows that the momentum does
not experience a corresponding increase of uncertainty. Is this a violation of
the uncertainty principle?

As a matter of fact, one should not be surprised with the experimental result
and should not consider this question as a new challenge. Similar results have
been demonstrated in EPR type of experiments and the same question has been
asked in EPR's 1935 paper \cite{EPR}.  In the past decades, we have been
worrying about problems concerning causality, locality, and reality more than
the ``crux" of the EPR paradox itself: the uncertainty principle.

\section{Popper's Experiment}

\begin{figure}[tbp]
\centerline{\epsfxsize=3in \epsffile{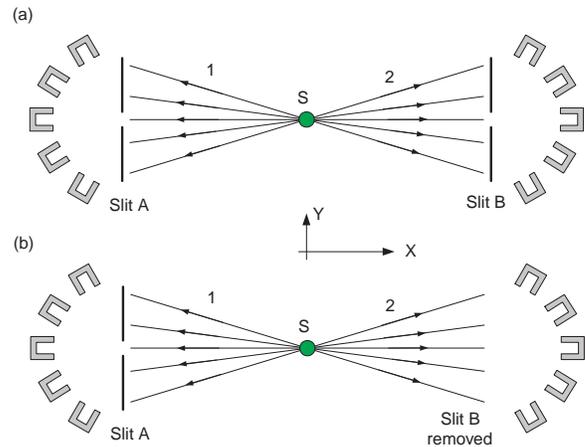}}
\caption{Popper's proposed experiment. An entangled pair of
particles are emitted from a point source with momentum
conservation. A narrow slit on screen A is placed in the path of
particle 1 to provide the precise knowledge of its position on
the $y$-axis and this also determines the precise $y$-position of
its twin, particle 2 on screen B. (a) Slits A and B are adjusted
both very narrowly. (b) Slit A is kept very narrow and slit B is
left wide open.} \label{Popperidea}
\end{figure}

Similar to the EPR's {\em gedankenexperiment}, Popper's experiment is also
based on the feature of {\em two-particle entanglement}. Quantum mechanics
allows the entangled EPR-type state, a state in which if the position or
momentum of particle 1 is known the corresponding observable of its twin,
particle 2, is then 100\% determined \cite{EPR}. Popper's original thought
experiment is schematically shown in Fig. \ref{Popperidea}. A point source S,
positronium as Popper suggests, is placed at the center of the experimental
arrangement from which entangled pairs of particles 1 and 2 are emitted in
opposite directions along the respective positive and negative $x$-axes towards
two screens A and B. There are slits on both screens parallel to the $y$-axis
and the slits may be adjusted by varying their widths $\Delta y$. Beyond the
slits on each side stand an array of Geiger counters for the coincidence
measurements of the particle pairs as shown in the figure. The entangled pair
could be emitted to any direction in $4\pi$ solid angles from the point source.
However, if particle 1 is detected in a certain direction then particle 2 is
known to be in the opposite direction due to the momentum conservation of the
quantum pair.

First, let us imagine the case in which slits A and B are adjusted both very
narrowly. In this circumstance, counters should come into play which are higher
up and lower down as viewed from the slits. The firing of these counters is
indicative of the greater $\Delta p_{y} $ due to the smaller $\Delta y$ for
each particle. There seems to be no disagreement in this situation between both
the Copenhagen school and Popper and both sides can provide a reasonable
explanation according to their own philosophical beliefs.

Next, suppose we keep the slit at A very narrow and leave the slit at B wide
open. The main purpose of the narrow slit A is to provide the precise knowledge
of the position $y$ of particle 1 and this subsequently determines the precise
position of its twin (particle 2) on side B through quantum entanglement. Now,
asks Popper, in the absence of the physical interaction with an actual slit,
does particle 2 experience a greater uncertainty in $\Delta p_{y}$ due to the
precise knowledge of its position? Based on his ``statistical-scatter" theory,
Popper provides a straightforward prediction: {\em particle 2 must not
experience a greater $\Delta p_{y}$ unless a real physical narrow slit B is
applied}. However, if Popper's conjecture is correct, this would imply the
product of $\Delta y$ and $\Delta p_{y}$ of particle 2 could be smaller than
$h$ ($\Delta y\Delta p_{y}< h $). This may pose a serious difficulty for the
Copenhagen camp and perhaps for many of us.  On the other hand, if particle 2
going to the right does scatter like its twin which has passed though slit A,
even though slit B is wide open, we are then confronted with an apparent {\it
action-at-a-distance}!

\section{Realization of Popper's Experiment}

\begin{figure}[tbp]
\centerline{\epsfxsize=3in \epsffile{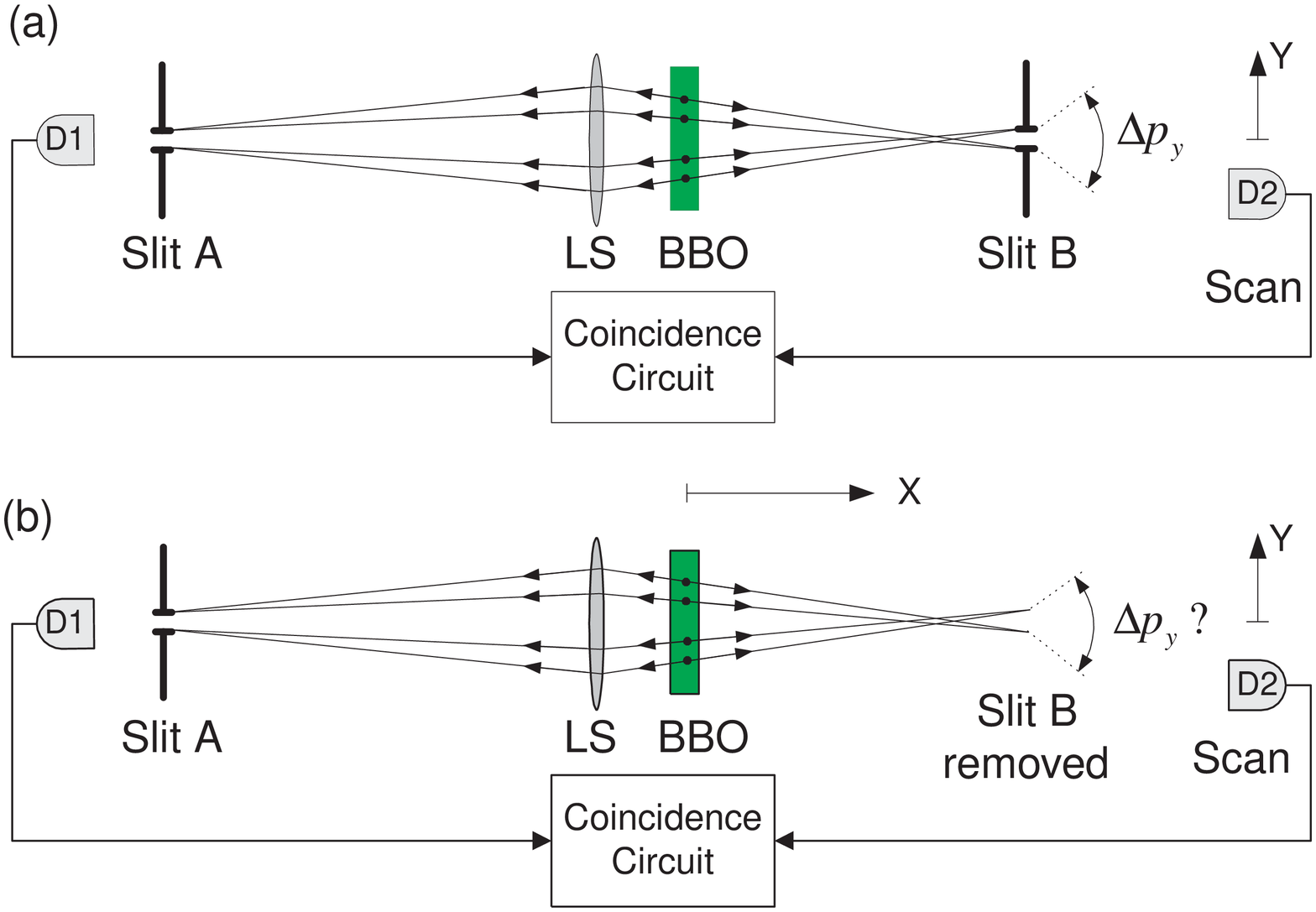}}
\caption{Modified version of Popper's experiment. An EPR photon
pair is generated by SPDC. A lens and a narrow slit A are placed
in the path of photon 1 to provide the precise knowledge of its
position on the $y$-axis and also determines the precise
$y$-position of its twin, photon 2, on screen B due to a ``ghost
image" effect. Two detectors $D_{1}$ and $D_{2}$ are used to scan
in the $y$-directions for coincidence counts. (a) Slits A and B
are adjusted both very narrowly. (b) Slit A is kept very narrow
and slit B is left wide open.} \label{PopperSPDC}
\end{figure}

We have realized Popper's experiment with the use of the entangled two-photon
source of spontaneous parametric down conversion (SPDC)\cite{Klyshko} \cite
{Yariv}. In order to clearly demonstrate all aspects of the historical and
modern experimental concerns in a practical manner, Popper's original design is
slightly modified as shown in Fig. \ref{PopperSPDC}. The two-photon source is a
CW Argon ion laser pumped SPDC which provides a two-photon entangled state that
preserves momentum conservation for the signal-idler photon pair in the SPDC
process. By taking advantage of the nature of entanglement of the signal-idler
pair (also labeled ``photon 1" and ``photon 2") one could make a ``ghost image"
of slit A at ``screen" B, see Fig. \ref{Popperunfold}. The physical principle
of the two-photon ``ghost image" has been reported in Ref. \cite{Todd}.

The experimental condition specified in Popper's experiment is then achieved:
when slit A is adjusted to a certain narrow width and slit B is wide open, slit
A provides precise knowledge about position of photon 1 on the $y$-axis up to
an accuracy $\Delta y$ which equals the width of slit A and the corresponding
``ghost image" of pinhole A at ``screen" B determines the precise position $y$
of photon 2 to within the same accuracy $\Delta y$. $ \Delta p_{y}$ of ``photon
2" can be independently studied by measuring the width of its ``diffraction
pattern'' at a certain distance from ``screen" B. This is obtained by recording
coincidences between detectors $D_{1}$ and $ D_{2}$ while scanning detector
$D_{2}$ along its $y$-axis which is behind ``screen" B at a certain distance.
Instead of a battery of Geiger counters, in our experiment only two photon
counting detectors $D_{1}$ and $D_{2}$ placed behind the respective slits A and
B are used for the coincidence detection. Both $D_{1}$ and $D_{2}$ are driven
by step motors and so can be scanned along their $y$-axes. $\Delta y\Delta
p_{y}$ of ``photon 2'' is then readily calculated and compared with $h $
\cite{FeynmanLecture}.

The use of a ``point source" in the original proposal has been much criticized
and considered as the fundamental mistake Popper made \cite{pointsource}
\cite{Horne}. The major objection is that a point source can never produce a
pair of entangled particles which preserves two-particle momentum conservation.
However, notice that a ``point source" is {\it not} a necessary requirement for
Popper's experiment. What is required is the position entanglement of the
two-particle system: if the position of particle 1 is precisely known, the
position of particle 2 is also 100\% determined. So one can learn the precise
knowledge of a particle's position through quantum entanglement. Quantum
mechanics does allow the position entanglement for an entangled system (EPR
state) and there are certain practical mechanisms, such as that the
``ghost-image" effect shown in our experiment, that can be used for its
realization.

\begin{figure}[tbp]
\centerline{\epsfxsize=3in \epsffile{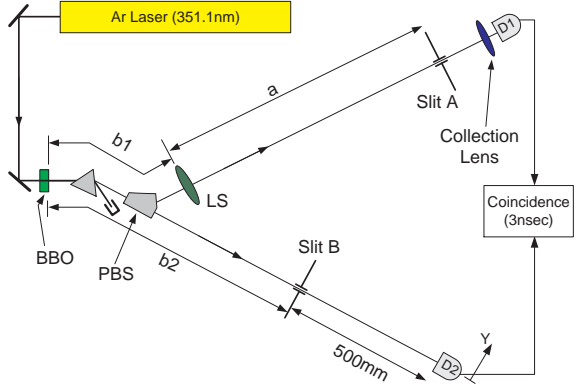}}
\caption{Schematic of the experimental setup. The laser beam is
about $3mm$ in diameter. The ``phase-matching condition" is well
reinforced. Slit A ($ 0.16mm$) is placed $1000mm=2f$ behind the
converging lens, LS ($f=500mm$). The one-to-one ``ghost image"
($0.16mm$) of slit A is located at B. The optical distance from
LS in the signal beam taken as back through PBS to the SPDC
crystal ($b_1=255mm$) and then along the idler beam to ``screen
B" ($b_2=745mm$) is $1000mm=2f$ ($b=b_1+b_2$). }
\label{Poppersetup}
\end{figure}

The schematic experimental setup is shown in Fig.\ref{Poppersetup} with
detailed indications of the various distances. A CW Argon ion laser line of
$\lambda _{p}=351.1nm$ is used to pump a $3mm$ long beta barium borate (BBO)
crystal for type-II SPDC \cite{type} to generate an orthogonally polarized
signal-idler photon pair. The laser beam is about $3mm$ in diameter with a
diffraction limited divergence. It is important not to focus the pump beam so
that the phase-matching condition, $ {\bf k}_{s}+{\bf k}_{i}={\bf k}_{p}$, is
well reinforced in the SPDC process \cite{Yariv}, where ${\bf k}_{j}$
$(j=s,i,p)$ is the wavevectors of the signal (s), idler (i), and pump (p)
respectively. The collinear signal-idler beams, with $\lambda _{s}=\lambda
_{i}=702.2nm=2\lambda _{p}$ are separated from the pump beam by a fused quartz
dispersion prism, and then split by a polarization beam splitter PBS. The
signal beam (``photon 1") passes through the converging lens LS with a $500mm$
focal length and a $25mm$ diameter. A $ 0.16mm$ slit is placed at location A
which is $1000mm$ $(=2f)$ behind the lens LS. The use of LS is to achieve a
``ghost image" of slit A ($0.16mm$) at ``screen" B which is at the same optical
distance $1000mm$ $(=2f)$ from LS, however in the idler beam (in the path of
``photon 2"). The signal and idler beams are then allowed to pass through the
respective slits A and B (a real slit B and then a ``ghost image" of slit A)
and to trigger the two photon counting detectors $D_{1}$ and $D_{2}$. A short
focal length lens is used with $D_{1}$ for collecting the signal beam which
passes through slit A. The point-like detector $D_{2}$ is located $500mm$
behind ``screen" B. The detectors are Geiger mode avalanche photodiodes which
are $ 180\mu m$ in diameter. $10nm$ band-pass spectral filters centered at
$702nm$ are used with each of the detectors. The output pulses from the
detectors are sent to a coincidence circuit. During the measurements, detector
$D_{1}$ is fixed behind slit A while detector $D_{2}$ is scanned on the
$y$-axis by a step motor.

{\bf Measurement 1}: we first studied the case in which both slits A and B were
adjusted to be $0.16mm$. The $y$-coordinate of $D_{1}$ was chosen to be $0$
(center) while $D_{2}$ was allowed to scan along its $y$-axis. The circled dot
data points in Fig. \ref{Popperdata} show the {\em coincidence} counting rates
against the $y$-coordinates of $D_{2}$. It is a typical single-slit diffraction
pattern with $\Delta y\Delta p_{y}=h $. Nothing is special in this measurement
except we have learned the width of the diffraction pattern for the $0.16mm$
slit and this represents the minimum uncertainty of $ \Delta p_{y}$
\cite{FeynmanLecture}. We should remark at this point that the {\em single}
detector counting rates of $D_{2}$ is basically the same as that of the
coincidence counts except for a higher counting rate.

{\bf Measurement 2}: the same experimental conditions were maintained except
that slit B was left wide open. This measurement is a test of Popper's
prediction. The $y $-coordinate of $D_{1}$ was chosen to be $0$ (center) while
$D_{2}$ was allowed to scan along its $y$-axis. Because of entanglement of the
signal-idler photon pair and the coincidence measurement, only those twins
which have passed through slit A and the ``ghost image" of slit A at ``screen"
B with an uncertainty of $\Delta y=0.16mm$ (which is the same width as the real
slit B we have used in measurement 1) would contribute to the coincidence
counts through the simultaneous triggering of $D_{1}$ and $D_{2}$. The diamond
dot data points in Fig. \ref{Popperdata} report the measured coincidence
counting rates against the $y$ coordinates of $D_{2}$. The measured width of
the pattern is narrower than that of the diffraction pattern shown in
measurement 1. At the same time, the width of the pattern is found to be much
narrower than the actual size of the diverging SPDC beam at $D_{2}$.  It is
also interesting to notice that the single counting rate of $D_{2}$ keeps
constant in the entire scanning range, which is very different from that in
measurement 1. The experimental data has provided a clear indication of $\Delta
y\Delta p_{y}< h$ in the coincidence measurements.

\section{Quantum Mechanical Prediction}

Given that $\Delta y\Delta p_{y}< h$, is this a violation of
uncertainty principle?  Before drawing any conclusion, let us
first examine what quantum mechanics predicts. If quantum
mechanics does provide a solution with $\Delta y\Delta p_{y}< h $
for ``photon 2". Indeed, we would be forced to face a paradox as
EPR had pointed out in 1935.

We begin with the question: how does one learn the precise position knowledge
of photon 2 at ``screen" B quantum mechanically? Is it really that $0.16mm$ as
determined by the width of slit A? The answer is in the positive. Quantum
mechanics predicts a ``ghost" image of slit A at ``screen" B which is $0.16mm$
for the above experimental setup. The crucial point is we are dealing with an
entangled two-photon state of SPDC \cite{Klyshko}\cite{Rubin},
\begin{eqnarray}
\left| \Psi \right\rangle  = \sum_{s,i}\delta \left( \omega
_{s}+\omega _{i}-\omega _{p}\right) \delta \left( {\bf
k}_{s}+{\bf k}_{i}-{\bf k} _{p}\right) \nonumber\\ \times
a_{s}^{\dagger }(\omega ({\bf k}_{s}))\ a_{i}^{\dagger }(\omega
({\bf k} _{i}))\left| 0\right\rangle , \label{state}
\end{eqnarray}
where $\omega _{j}$, {\bf k$_{j}$ (}j = s, i, p) are the frequencies and
wavevectors of the signal (s), idler (i), and pump (p) respectively. $\omega
_{p}$ and {\bf k}$_{p}$ can be considered as constants while $a_{s}^{\dagger }$
and $a_{i}^{\dagger }$ are the respective creation operators for the signal and
the idler. As given in the above form, the entanglement feature in state (1)
may be thought of as the superposition of an infinite number of ``two-photon''
states that corresponds to the infinite numbers of ways the SPDC signal-idler
can satisfy the conditions of energy and momentum conservation, as represented
by the $ \delta $-functions of the state which is technically known as
phase-matching conditions:
\begin{equation}
\omega _{s}+\omega _{i}=\omega _{p},\quad {\bf k}_{s}{\bf
+k}_{i}={\bf k}_{p} .\label{phasematch}
\end{equation}
It is interesting to see that even though there is no precise knowledge of the
momentum for either the signal or the idler, the state nonetheless provides
precise knowledge of the {\em momentum correlation} of the pair. In the
language of EPR, the momentum for neither the signal photon nor the idler
photon is determined but if a measurement on one of the photons yields a
certain value, the momentum of the other photon is 100\% determined.

\begin{figure}[tbp]
\centerline{\epsfxsize=3in \epsffile{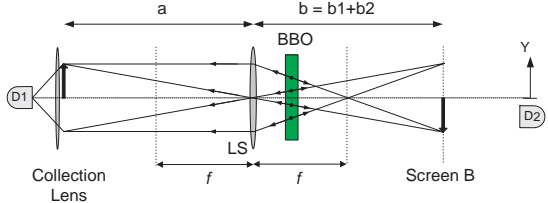}}
\caption{The unfolded schematic of the experiment. This is
equivalent to assume ${\bf k}_{s}+{\bf k}_{i}=0$ but without
losing the important entanglement feature of the momentum
conservation of the signal-idler pair. It is clear that the
locations of slit A, lens LS, and the ``ghost image" must be
governed by the Gaussian thin lens equation, bearing in mind the
different propagation directions of the signal-idler by the small
arrows on the straight-line two-photon paths.}
\label{Popperunfold}
\end{figure}

To simplify the physical picture, we ``unfold" the signal-idler
paths in the schematic of Fig.\ref{Poppersetup} into that shown in
Fig.\ref{Popperunfold}, which is equivalent to assuming ${\bf
k}_{s}+{\bf k}_{i}=0$ while not losing the important entanglement
feature of the momentum conservation of the signal-idler pair.
This important peculiarity selects the only possible optical paths
of the signal-idler pairs that result in a ``click-click"
coincidence detection which are represented by {\em straight
lines} in this unfolded version of the experimental schematic so
that the ``image" of slit A is well-produced in {\em coincidences}
as shown in the figure. It is similar to an optical imaging in the
``usual" geometric optics picture, bearing in mind the different
propagation directions of the signal-idler indicated by the small
arrows on the {\em straight lines}. It is easy to see that a
``clear" image requires the locations of slit A, lens LS, and
screen B to be governed by the Gaussian thin lens equation
\cite{Todd},
\begin{equation}
\frac{1}{a}+\frac{1}{b}=\frac{1}{f}.  \label{Gaussian}
\end{equation}
In our experiment, we have chosen $a=b=2f=1000mm$, so that the ``ghost image"
of slit A at ``screen" B must have the same width as that of slit A.  The
measured size of the ``ghost image" agrees with theory.

In Fig. \ref{Popperunfold} we see clearly these two-photon paths ({\em straight
lines}) that result in a ``click-click" joint detection are restricted by slit
A, lens LS as well as momentum conservation. As a result, any signal-idler pair
that passes through the $0.16mm$ slit A would be ``localized" within $\Delta
y=0.16mm$ at ``screen" B. In this way, one does learn the precise position
knowledge of photon 2 through the entanglement nature of the two-photon system.

One could also explain this ``ghost image" in terms of conditional
measurements: conditioned on the detection of ``photon 1" by detector $D_1$
behind slit A, ``photon 2" can only be found in a certain position. In other
words, ``photon 2" is localized only upon the detection of photon 1.

Now let us go further to examine $\Delta p_{y}$ of photon 2 which is
conditionally ``localized" within $\Delta y=0.16mm$ at ``screen" B. In order to
study $\Delta p_{y}$, the photon counting detector $D_2$ is scanned $ 500mm$
behind ``screen" B to measure the ``diffraction pattern".  $\Delta p_{y}$ can
be easily estimated from the measurement of the width of the diffraction
pattern \cite{FeynmanLecture}. The two-photon paths, indicated by the {\em
straight lines} reach detector 2 which is located $500mm$ behind ``screen" B so
that detector $D_{2}$ will receive ``photon 2" in a much narrower width under
the condition of the ``click" of detector $D_1$ as shown in measurement 2,
unless a real physical slit B is applied to ``disturb" the {\em straight
lines}.

Apparently we have a paradox: quantum mechanics provides us with a solution
which gives $\Delta y\Delta p_{y}<h $ in measurement 2 and the experimental
measurements agree with the prediction of quantum mechanics.

\section{Conclusion}

It is the same paradox of EPR.  Indeed, one could consider this experiment as a
variant of the 1935 EPR {\em gedankenexperiment} in which the position-momentum
uncertainty was questioned by Einstein-Podolsky-Rosen based on the discussion
of a two-particle entangled state \cite{EPR}. Comparing with the EPR-Bohm
experiment \cite{Bohm}, which is a simplified version of the 1935 EPR {\em
gedankenexperiment}, the spin for neither particle is determined (uncertain);
however, if one particle is measured to be spin up along a certain direction,
the other one must be spin down along that direction (certain). All the spin
components of a particle can be precisely determined through the measurement of
its twin.

Quantum mechanics gives prediction for the EPR and the EPR-Bohm correlations
based on the measurements for entangled states. All reported historical
experiments have shown good agreement with quantum mechanics as well as EPR's
prediction (but not their interpretation). The results of our experiment agree
with quantum mechanics and Popper's prediction too.  We therefore consider the
following discussions may apply to both EPR and Popper.

Popper and EPR were correct in the prediction of the physical outcomes of their
experiments. However, Popper and EPR made the same error by applying the
results of two-particle physics to the explanation of the behavior of an
individual particle. The two-particle entangled state is not the state of two
individual particles. Our experimental result is emphatically NOT a violation
of the uncertainty principle which governs the behavior of an individual
quantum.

In both the Popper and EPR experiments the measurements are ``joint detection"
between two detectors applied to entangled states. Quantum mechanically, an
entangled two-particle state only provides {\em the precise knowledge of the
correlations of the pair}. Neither of the subsystems is determined by the
state. It can be clearly seen from our above analysis of Popper's experiment
that this kind of measurements is only useful to decide on how good the
correlation is between the entangled pair. In other words, the behavior of
``photon 2'' observed in our experiment is conditioned upon the measurement of
its twin.  A quantum must obey the uncertainty principle but the ``conditional
behavior" of a quantum in an entangled two-particle system is different. The
uncertainty principle is not for ``conditional'' behavior.  We believe
paradoxes are unavoidable if one insists the {\em conditional behavior} of a
particle is the {\em behavior} of a particle. This is the central problem of
the rationale behind both Popper and EPR.  $\Delta y\Delta p_{y}\geq h$ is not
applicable to the conditional behavior of either ``photon 1" or ``photon 2" in
the case of the Popper and EPR type of measurements.

\begin{figure}[tbp]
\centerline{\epsfxsize=3in \epsffile{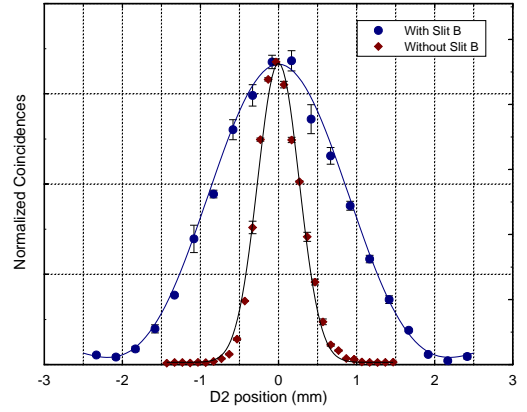}}
\caption{The observed coincidence patterns. The $y$-coordinate of
$D_{1}$ was chosen to be $0$ (center) while $D_{2}$ was allowed
to scan along its $y$-axis. Circled dot points: {\em Slit A =
Slit B = $0.16mm$}. Diamond dot points: {\em Slit A = $0.16mm$,
Slit B wide open}. The width of the $sinc$ function curve fitted
by the circled dot points is a measure of the minimum $ \Delta
p_{y}$ determined by a $0.16mm$ slit. } \label{Popperdata}
\end{figure}

The behavior of photon 2 conditioned upon photon 1 is well represented by the
two-photon amplitudes.  Each of the {\em straight lines} in the above
discussion corresponds to a two-photon amplitude.  Quantum mechanically, the
superposition of these two-photon amplitudes are responsible for
a``click-click" measurement of the entangled pair.  A ``click-click'' joint
measurement of the two-particle entangled state projects out certain
two-particle amplitudes and only these two-particle amplitudes feature in the
quantum formalism.  In the above analysis we never consider ``photon 1'' or
``photon 2'' {\em individually}.  Popper's question about the momentum
uncertainty of photon 2 is then inappropriate. The correct question to ask in
these measurements should be: what is the $\Delta p_{y}$ for the signal-idler
pair which are ``localized" within $\Delta y=0.16mm$ at ``screen" B and at
``screen" A and governed by the momentum conservation? This is indeed the
central point for this experiment. There is no reason to expect
the``conditionally localized photon 2" will follow the familiar interpretation
of the uncertainty relation as shown in Fig. \ref{Popperdata}.

\begin{figure}[tbp]
\centerline{\epsfxsize=3in \epsffile{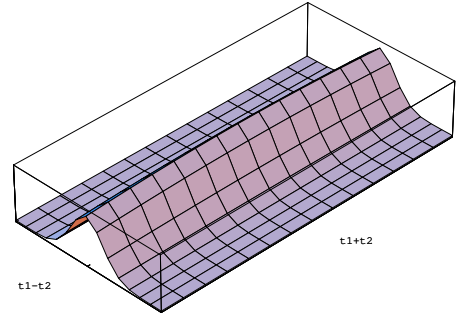}}
\caption{Biphoton wavepacket envelope calculated from the state
of type-I spontaneous parametric down conversion.  For a
simplified situation, it can be written as:
$\Psi(t_{1},t_{2})=A_{0}e^{-\sigma_{+}^{2}(t_{1}+t_{2})^{2}}
e^{-\sigma_{-}^{2}(t_{1}-t_{2})^{2}}e^{-i\Omega_{s}t_{1}}e^{-i\Omega
_{i}t_{2}}$, where $\Omega _{j}$, $j=s,i$, is the central
frequency for signal or idler, $1/\sigma _{\pm }$ are coherence
times, $t_{i}\equiv T_{i}-L_{i}/c,\;i=1,2$, $T_{i}$ is the
detection time of detector $i$ and $L_{i}$ the optical pathlength
of the signal or idler from SPDC to the $i$th detector.
$\Psi(t_{1},t_{2})$ is a non-factorizable two-dimensional
wavepacket, we may call it {\em biphoton}.}\label{Biphoton}
\end{figure}

Quantum mechanics shows that the superposition of these
two-photon amplitudes results in a non-factorizable
two-dimensional {\em biphoton} wavepacket
\cite{schro}\cite{packet}\cite{Rubin} instead of two individual
wavepackets associated with photon 1 and photon 2. Figure
\ref{Biphoton} gives a simple picture of the {\em biphoton}
wavepacket of SPDC.  We believe all the problems raised by the
EPR and Popper type experiments can be duly resolved if the
concept of {\em biphoton} is adopted in place of two individual
photons.

Once again, this recent demonstration of the thought experiment of Popper calls
our attention to the important message: the physics of the entangled
two-particle system must inherently be very different from that of individual
particles.  In the spirit of the above discussions, we conclude that it has
been a long-standing historical mistake to mix up the uncertainty relations
governing an individual single particle with an entangled two-particle system.

\vspace{1cm}

The authors acknowledge important suggestions and encouragement
from T. Angelidis, A. Garuccio, C.K.W. Ma, and J.P. Vigier. We
especially thank C.K.W. Ma from LSE for many helpful discussions.
We are grateful to A. Sudbery for useful comments. This research
was partially supported by the U.S. Office of Naval Research and
the U.S. Army Research Office - National Security Agency grants.

\begin{references}

\bibitem{Popper1}  K.R. Popper, `Zur Kritik der
Ungenauigkeitsrelationen' , {\em Die Naturwissenschaften}, 22,
807 (1934); K.R. Popper,{\em Quantum Theory and the Schism in
Physics} (Hutchinson, London, 1983). Amongst the most notable
opponents to the ``Copenhagen School" were
Einstein-Podolsky-Rosen, de Broglie, Land\'{e}, and Karl Popper.
One may not agree with Popper's philosophy (EPR classical reality
as well) but once again, Popper's thought experiment brings yet
attention to the fundamental problems of quantum theory.

\bibitem{EPR}  A. Einstein, B. Podolsky, and N. Rosen, {\em Phys. Rev.},
{\bf 47}, 777 (1935).

\bibitem{Klyshko}  D.N. Klyshko, {\em Photon and Nonlinear Optics} (Gordon and Breach
Science, New York, 1988).

\bibitem{Yariv}  A. Yariv, {\em Quantum Electronics}, (John Wiley and Sons, New York, 1989).

\bibitem{Todd}  T.B. Pittman, Y.H. Shih, D.V. Strekalov, and A.V. Sergienko,
{\em Phys. Rev. A}, {\bf 52}, R3429 (1995).

\bibitem{FeynmanLecture} R.P. Feynman, {\em The Feynman Lectures on Physics},
Vol. III, (Addison-Wesley, Reading, Massachusetts, 1965).

\bibitem{pointsource}  For criticisms of Popper's experiment, see for example,
D. Bedford and F. Selleri, {\em Lettere al Nuovo Cimento}, {\bf
42}, 325 (1985); M.J. Collett and R. Loudon, {\em Nature}, {\bf
326}, 671 (1987); A. Sudbery, {\em Philosophy of Science}, {\bf
52}, 470 (1985); A. Sudbery, {\em Microphysical Reality and
Quantum Formalism}, A. van der Merwe et al. (eds.) (Kluwer
Academic, Dordrecht, 1988). Many of the criticisms concern the
validity of a point source for entangled two-particles. However,
a ``point source'' is not a necessary requirement for Popper's
experiment. What is essential is to learn the precise knowledge
of a particle's position through quantum entanglement. This is
achieved in our experiment by means of a ``ghost image"
\cite{Todd}.

\bibitem{Horne} For discussions of the effect of the size of the source
on one-particle and two-particle diffraction, see for example, M. Horne, {\em
Experimental Metaphysics}, eds. R.S. Cohen, M. Horne, and J. Stachel (Kluwer
Academic, Dordrecht, 1997).

\bibitem{Rubin} M.H. Rubin, D.N. Klyshko, and Y.H. Shih, {\em Phys. Rev. A}
{\bf 50}, 5122 (1994).

\bibitem{type}  In type-I SPDC, signal and idler are both ordinary (or
extraordinary) rays of the crystal; however, in type-II SPDC they are
orthogonally polarized, i.e., one is ordinary and the other is extraordinary.

\bibitem{Bohm}  D. Bohm, {\em Quantum Theory} (Prentice Hall Inc., New York, 1951).

\bibitem{schro}  E.Schr\"{o}dinger, {\em Naturwissenschaften}, {\bf 23}, 807, 823,
844 (1935); translations appear in {\em Quantum Theory and
Measurement}, eds. J.A. Wheeler and W.H. Zurek (Princeton
University Press, New York, 1983).

\bibitem{packet}  Y.H. Shih and A.V. Sergienko, {\em Phys. Rev. A} {\bf 50}, 2564
(1994); A.V. Sergienko, Y.H. Shih, and M.H. Rubin, {\em J. Opt.
Soc. Am. B} {\bf 12}, 859 (1995).

\end{references}
\end{document}